\documentclass{vgtc}                          

\ifpdf
  \pdfoutput=1\relax                   
  \usepackage{graphicx}                
  \DeclareGraphicsExtensions{.pdf,.png,.jpg,.jpeg} 
\else
  \usepackage{graphicx}                
  \DeclareGraphicsExtensions{.eps}     
\fi%

\graphicspath{{figures/}{pictures/}{images/}{./}} 

\usepackage{microtype}                 
\PassOptionsToPackage{warn}{textcomp}  
\usepackage{textcomp}                  
\usepackage{mathptmx}                  
\usepackage{times}                     
\usepackage{cite}                      
\usepackage{tabu}                      
\usepackage{booktabs}                  
\usepackage{hyperref}
\usepackage{cleveref}

\usepackage{color}

\onlineid{1070}

\vgtccategory{Research}

\vgtcinsertpkg

\title{Understanding Auto-Scheduling Optimizations for Model Deployment via Visualizations}

 \author{Laixin Xie\thanks{e-mail: xielx@shanghaitech.edu.cn}\\ %
         \scriptsize ShanghaiTech University, Shanghai%
 \and Chenyang Zhang\thanks{e-mail: zhang414@illinois.edu}\\ %
      \scriptsize University of Illinois at Urbana-Champaign %
      \and Ruofei Ma\thanks{e-mail: marf@shanghaitech.edu.cn}\\ %
      \scriptsize ShanghaiTech University, Shanghai %
 \and Xing Jiang \thanks{e-mail: jiangxing@corp.netease.com}\\ %
  \scriptsize UX Center, Netease Games, Shanghai %
 \and Xingxing Xing \thanks{e-mail: xingxingxing@corp.netease.com}\\
  \scriptsize UX Center, Netease Games, Guangzhou %
 \and Wei Wan \thanks{e-mail: gzwanwei@corp.netease.com}\\
  \scriptsize UX Center, Netease Games, Hangzhou %
 \and Quan Li\thanks{e-mail: liquan@shanghaitech.edu.cn (corresponding author)}\\ %
     \parbox{4in}{\scriptsize \centering School of Information Science and Technology, ShanghaiTech University, Shanghai\\Shanghai Engineering Research Center of Intelligent Vision and Imaging, China}
 }

\teaser{
  \centering
    \vspace{-4mm}
  \includegraphics[width=\linewidth]{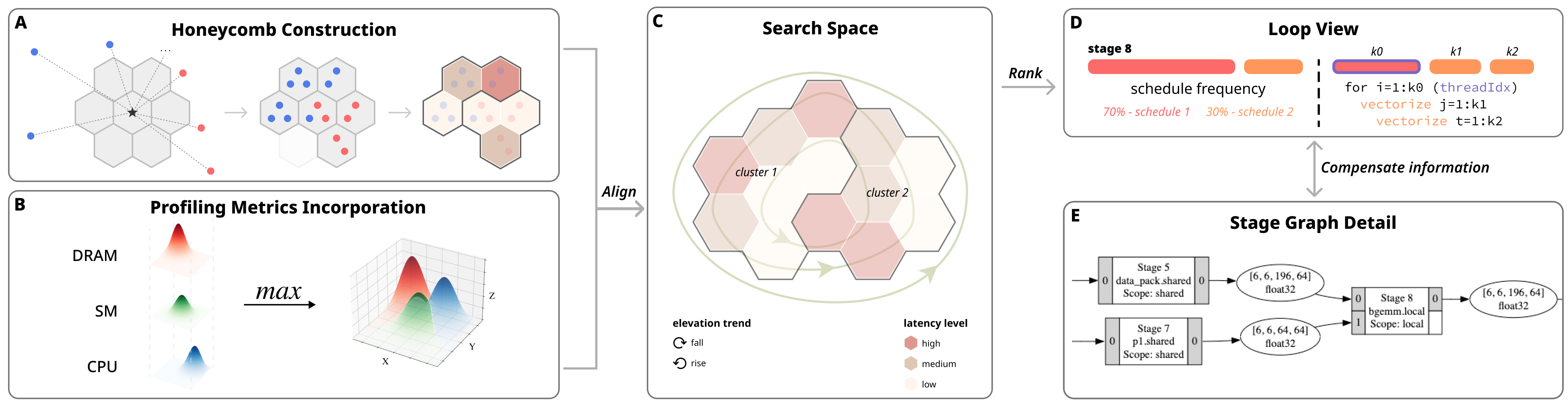}
  \vspace{-6mm}
  \caption{%
  	(A) Honeycomb Chart Construction; (B) Hardware Performance Comparison; (C) Honeycomb Chart and Contour Map in Search Space Visualization; (D) Example Entry in the Loop View; (E) Stage Graph visualization by TEDD~\cite{TEDD}.
  }
  \label{fig:teaser}
      \vspace{-3mm}
}

\abstract{After completing the design and training phases, deploying a deep learning model onto specific hardware is essential before practical implementation. Targeted optimizations are necessary to enhance the model's performance by reducing inference latency. Auto-scheduling, an automated technique offering various optimization options, proves to be a viable solution for large-scale auto-deployment. However, the low-level code generated by auto-scheduling resembles hardware coding, potentially hindering human comprehension and impeding manual optimization efforts. In this ongoing study, we aim to develop an enhanced visualization that effectively addresses the extensive profiling metrics associated with auto-scheduling. This visualization will illuminate the intricate scheduling process, enabling further advancements in latency optimization through insights derived from the schedule.}

\CCScatlist{
  \CCScatTwelve{Program Analysis}{Compilation}{Visualization}{Model Deployment}
}

\begin{document}


\firstsection{Introduction}

\maketitle

\par Model deployment is a crucial stage in utilizing trained models practically, following training and validation. During this phase, models transition from research to production environments tailored to specific hardware. Beyond model accuracy, considerations encompass usability and reliability. Of these, inference latency in deployed models emerges as a critical concern, particularly in contexts like autonomous driving where high latency may lead to accidents. To achieve low latency while preserving model output integrity, deployment engineers fine-tune the program's running schedule, which governs the mapping of tensor operations to hardware resources. However, this task is time-intensive and requires substantial expertise, especially for large-scale deployments across diverse platforms. Thus, the development of a scalable and interpretable tool to support engineers in efficiently monitoring, comprehending, and optimizing inference latencies during deployment is a significant challenge.


\par Two approaches have emerged to enhance model deployers' effectiveness. The first approach combines existing visualizations with program profiling techniques, enabling deployers to understand the model's dynamic behavior during runtime, identify bottlenecks, and detect anomalies. However, these approaches lack specific guidance on optimal solutions, requiring significant manual effort from deployers to implement low-level code based on their expertise. The second approach involves automatic code generation through auto-scheduling by the compiler, leveraging hardware characteristics to optimize execution performance~\cite{zheng2020ansor}. While it reduces the coding burden for deployers, the lack of interpretability can lead to inconsistent performance improvements and confusion regarding underlying mechanisms. To address these challenges, we propose a hybrid approach that combines profiling and auto-scheduling techniques to harness the advantages of both methodologies. In this study, we leverage the extensive range of optimization options provided by Ansor~\cite{zheng2020ansor} to develop two distinct visual designs. The first visualization assists deployers in selecting relevant information from profiling metrics, enabling informed decision-making for latency optimization. The second visualization transforms complex code loops associated with the schedule into an intuitive visual representation, enhancing deployers' comprehension and facilitating further latency optimization efforts using insights derived from the schedule.

\begin{figure*}[htp] 
\vspace{-4mm}
  \centering 
  \includegraphics[width=\textwidth]{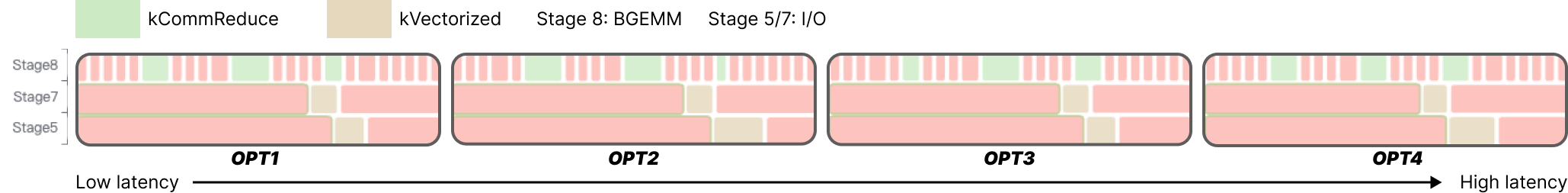} 
  \vspace{-6mm}
  \caption{Deriving design principles for parallelism techniques through visual analysis in model optimization.} 
  \label{fig:case} 
  \vspace{-6mm}
\end{figure*}

\section{Visual Designs}
\par We propose two visual designs for hierarchical analysis of auto-scheduling results. The first design focuses on exploring the search space documented in Ansor's searching log, which captures the search process over a computational graph with coarse operators like convolution, encompassing operations such as addition and multiplication. The second design, known as the ``loop view'', presents a visually intuitive schedule where each schedule corresponds to a fine-grained operation like BGEMM (Batched General Matrix Multiply), typically involving a single hardware instruction known as the ``stage graph''. Our visualizations break down the task graph into subtasks and transform them into stages, facilitating meticulous optimization of the computational graph.

\subsection{Search Space Visualization}
\par After identifying the task graph with the highest latency, a comparative analysis of performance improvements achieved through different auto-scheduling strategies can be conducted. To accomplish this, we \textit{construct an optimization search space} specifically for optimization, represented using a honeycomb design as the foreground. Additionally, we \textit{incorporate hardware profiling metrics} into our visualization as the background, aiding users in understanding the performance differences among various optimization options.




\par \textbf{Constructing the Optimization Search Space.} Drawing inspiration from Ansor, we extract distinctive features from optimized tensor programs to represent available optimization options, including arithmetic operations, loop transformations, and GPU thread binding. Employing dimension reduction and clustering techniques, we analyze and compare similarities among the optimization options. To visualize these options in a two-dimensional space while preserving their similarity, we introduce the ``honeycomb chart''. The chart is constructed step-by-step, with cells formed outward from the clustering center, prioritizing points closer to the core (\autoref{fig:teaser}(A)). Further partitioning of the honeycomb enhances understanding by assigning each cell a corresponding category based on the majority category within it. Boundaries are established between cells of different categories. Additionally, we compute the average inference latency for optimized options within each cell and represent it using a three-color scheme. Darker hues indicate higher average latencies, allowing users to identify clusters with exceptional or anomalous performance characteristics easily (\autoref{fig:teaser}(C)). This visual representation amalgamates similar options, organizes significant optimizations hierarchically, and ensures scalability.

\par \textbf{Incorporating Hardware Profiling Metrics.} The performance enhancement achieved by each optimization option is compared to the baseline (i.e., metrics without any optimization) across all profiling metrics. These metrics are categorized into three groups based on the corresponding hardware components: Dynamic Random Access Memory (DRAM), Streaming Multiprocessor (SM), and Graphics Processing Unit (GPU). To improve the coherence and visualization of performance variations in the honeycomb graph, we use honeycomb cells as the fundamental units. Within each cell, we calculate the average performance enhancement separately for the three hardware types. Next, we merge the distinct peaks by selecting the maximum value among them as the Z-axis value for each point in the XY plane (\autoref{fig:teaser}(B)). This merging process indicates the hardware type with the most substantial performance improvement due to associated optimization options within the honeycomb space. The resulting three-dimensional peaks are transformed into contour lines within the two-dimensional honeycomb space. The contour lines incorporate directional information, with a clockwise direction signifying a decrease towards the center and a counterclockwise direction indicating an increase away from the center, depicting the trend of peak heights. The contour are laid on the background of Honeycomb to elucidate hardware-oriented principles, enabling identification of optimal options and valuable insights for hardware utilization. It facilitates identifying diverse optimization patterns and learning from Ansor's searching logs.

\subsection{Understanding Optimization Option}
\par The ``Loop View'' consists of two components to explore scheduling strategies within each stage of the fine-grained computational graph. The left section shows the distribution of scheduling strategies based on frequency, while the right section reveals the proportion of loops with distinct scheduling strategies. For example, in ``stage 8'', two scheduling strategies are employed, with 70\% using type 1 and 30\% using type 2 (\autoref{fig:teaser}(D)). The number of loops with each strategy is indicated. The ``Stage Graph Detail'' (\autoref{fig:teaser}(E)) dissects data flow between different stages within the selected task, providing users with rich contextual information for evaluating optimization strategies for each stage in the scheduling.

\section{Evaluation}
\par In a case study with a domain expert, we utilized the ResNet-18 model. Through our visual designs, the expert explored and analyzed the optimization options, leading to the derivation of two design principles for the parameter of two parallelism techniques: \textit{commutative reducer} and \textit{vectorization}. \textbf{1) Shorten the commutative reducer in stage 8:} The expert observed reduced latency in the BGEMM operation with a smaller commutative reducer (OP1-4 comparison in \autoref{fig:case}). Based on this, the expert proposed \textbf{design principle 1}, suggesting incremental iteration of the final commutative reducer's search range within BGEMM's loop. \textbf{2) Shorten the vectorization in stage 5:} Comparing OPT3 with OPT4 (\autoref{fig:case}), the expert identified the potential to reduce latency by shortening the vectorization in stage 5. A shorter vectorization call increases throughput, particularly beneficial for stage 5 with numerous small loops. Thus, \textbf{design principle 2} suggests exploring a narrow range of values to determine the optimal length of the vectorization operation in stage 5. In conclusion, the expert effectively derived two design principles using our visual designs, enabling quick identification of optimal values and improved performance. This outcome provides strong evidence of the effectiveness of our visual designs in aiding the expert's analysis and decision-making process.

\section{Discussion and Future Work}
\par The deployment of deep learning models is vital for their widespread adoption. As Large Language Models (LLM) progress, achieving low inference latency becomes crucial for enhancing user experience by reducing waiting times for answers. Our contribution lies in the field of visualization for model deployment techniques. Our goal is to develop an analytical workflow and a visual analytics system with user feedback. We aim to evaluate the system's effectiveness in supporting deployers and reducing inference latency using quantitative and qualitative methods.


\bibliographystyle{abbrv-doi}

\bibliography{template}
\end{document}